\documentclass[
superscriptaddress,
preprint
]{revtex4}

\usepackage{graphicx}
\usepackage{dcolumn}
\usepackage{bm}
\usepackage{amsmath}
\usepackage{float}

\begin{document}

\title{Temperature Dependent Mean Free Path Spectra of Thermal Phonons Along the c-axis of Graphite}


\author{Hang Zhang}
\thanks{These authors contributed equally to this work.}

\affiliation{Division of Engineering and Applied Science\\
California Institute of Technology\\
Pasadena, CA 91125}

\author{Xiangwen Chen}
\thanks{These authors contributed equally to this work.}

\affiliation{Division of Engineering and Applied Science\\
California Institute of Technology\\
Pasadena, CA 91125}

\author{Young-Dahl Jho}
\affiliation{School of Information and Communications\\
Gwangju Institute of Science and Technology\\
Gwangju 500-712, Korea}

\author{Austin J. Minnich}
\email{aminnich@caltech.edu}
\affiliation{Division of Engineering and Applied Science\\
California Institute of Technology\\
Pasadena, CA 91125}


\begin{abstract}
Heat conduction in graphite has been studied for decades because of its exceptionally large thermal anisotropy. While the bulk thermal conductivities along the in-plane and cross-plane directions are well known, less understood are the microscopic properties of the thermal phonons responsible for heat conduction. In particular, recent experimental and computational works indicate that the average phonon mean free path (MFP) along the c-axis is considerably larger than that estimated by kinetic theory, but the distribution of MFPs remains unknown. Here, we report the first quantitative measurements of c-axis phonon MFP spectra in graphite at a variety of temperatures using time-domain thermoreflectance measurements of graphite flakes with variable thickness. Our results indicate that c-axis phonon MFPs have values of a few hundred nanometers at room temperature and a much narrower distribution than in isotropic crystals. At low temperatures, phonon scattering is dominated by grain boundaries separating crystalline regions of different rotational orientation. Our study provides important new insights into heat transport and phonon scattering mechanisms in graphite and other anisotropic van der Waals solids.\\
\end{abstract}

\maketitle


\pagebreak

Thermal transport in anisotropic layered materials has attracted broad interest in fundamental science and applications\cite{Minnich2015, Jang2013, Dresselhaus1987, Chiritescu2007,Yan2012, Losego2013, Schmidt2008, Luckyanova2013, Minnich2015a, Foley2015, Yang2014, Fu2015} due to their large thermal anisotropy. The best known of these materials, graphite, possesses ultrahigh in-plane thermal conductivity of $\sim$ 2000 W/mK\cite{Touloukian1970} and cross-plane (c-axis) thermal conductivity around 6.8 W/mK\cite{Taylor1966} at room temperature, yielding an anisotropy factor of around 300. During the past decade, research has mainly focused on thermal transport properties along the in-plane direction, especially in graphene, the ideal two-dimensional system isolated from the ab-plane\cite{Jang2010,Ghosh2010,Seol2010,Bonini2012,Lindsay2014,
Paulatto2013,Bae2013,Xu2014}. Such a high in-plane thermal conductivity has important applications in heat dissipation\cite{Yan2012,Ghosh2008}. In addition, the low thermal contact resistance of ultra-thin graphite and graphene flakes make them promising candidates for thermal interface materials\cite{Chen2009}.\\

Compared with works on the in-plane direction, thermal transport along the c-axis of graphite has not yet been extensively studied.  Early works used kinetic theory to estimate that the average MFP along the c-axis is only a few nanometers at room temperature\cite{Shen2013,Tanaka1972}. Recently, several studies have indicated that the average phonon mean free path along c-axis (c-MFP) is actually much higher than this simple estimation. Sadeghia et al. estimated the average c-MFP is about 20 nm\cite{Sadeghi2013}. Harb et al. reported a cross-plane thermal conductivity of only 0.7 W/mK in a 35 nm thick graphite film, giving strong evidence of boundary scattering of phonons with MFPs much longer than 35 nm\cite{Harb2012}. Using molecular dynamics simulations, Wei et al\cite{Wei2014}. observed the presence of very long c-MFP phonons on the order of several hundred nanometers. However, this study was limited in the thickest samples that could be studied by computational limitations. Experimental measurements by Yang et al. suggested that the average MFP along c-axis should be much greater than 92 nm at room temperature\cite{Yang2014}, but this estimation is based on a model of an isotropic van der Waals (vdW) solid rather than the actual anisotropic phonon dispersion\cite{Yang2014}. \\

A pioneering experimental work by Fu et al. has recently provided strong evidence that the average c-MFP at room temperature is around 204 nm\cite{Fu2015}. In this work, thickness dependent thermal conductivity of graphite ribbons was obtained with the 3$\omega$ method and the c-MFP estimated by fitting the experimental data\cite{Fu2015}. However, due to experimental limitations important questions still remain regarding the spectrum of c-MFPs over a range of temperatures. For example, the limited temperature range of the measurements of Fu et al makes identifying the dominant scattering mechanisms in different temperature regimes challenging. \\ 

In this work, we report the first quantitative measurements of the temperature-dependent c-MFP spectrum of graphite. Using a theoretical approach to interpret systematic measurements of c-axis thermal conductivity over a wide range of sample thicknesses and temperatures, we reconstruct the c-MFP spectrum at each temperature from which we can identify the key scattering mechanisms. We find that at room temperature, the c-MFPs range from 40 nm to 250 nm, a much narrower distribution than in isotropic crystalline materials but far larger than early estimates of a few nanometers. At low temperatures, the thermal conductivity is limited by phonon scattering at grain boundaries separating crystalline regions of different rotational orientations, yielding MFPs between 100-600 nm. Our study provides important new insights into the microscopic processes governing heat conduction in graphite. \\


We conducted thermal conductivity measurements of thin films of graphite using a thermal characterization technique, time-domain thermoreflectance (TDTR). TDTR is a standard characterization method in the thermal sciences that operates by impulsively heating a sample with a laser pulse and observing the transient thermal decay with a probe beam. The thermal conductivity is obtained by fitting the thermal decay curve to a thermal model \cite{Cahill2004,Schmidt2008}. In this study, we use a two-tint implementation of this method as described in Ref. \citenum{Kang2008}. Briefly, a Ti:sapphire laser oscillator generates a series of femtosecond pulses at a repetition frequency of 76 MHz and wavelength of 785 nm. The pump pulse impulsively heats the sample, and the time-delayed probe pulse measures the change in optical reflectance of the Al transducer due to the temperature change, yielding a temperature decay curve. The pump pulse train is modulated at a frequency between 0.5 MHz to 14 MHz to enable lock-in detection. The laser spot sizes of pump and probe ranged from 13.5 $\mu$m to 60 $\mu$m and 10 $\mu$m (fixed), respectively. The measurements were found to be independent on pump size. Low temperature measurements were performed in an optical cryostat under vacuum of around 10$^{-6}$ torr.\\

The samples in this study consisted of a sandwich structure as shown in Figure 1a, with the Al transducer film and sapphire substrate sandwiching graphite samples of various thicknesses. Graphite flakes were first mechanically exfoliated from bulk HOPG samples and placed on top of sapphire wafers, with adhesion being by the vdW force. Large, flat regions in graphite samples were carefully chosen via optical microscopy. Their thicknesses were characterized by atomic force microscope (AFM) (Supplementary Information). Then, the Al transducer layer was deposited on top of the graphite flakes by electron beam evaporation. The thicknesses of the Al films were measured by AFM and had typical values of 70 nm to 86 nm. Figure 1b shows an optical image of one typical sample in our experiment (see additional images in the SI). Samples were carefully selected from a large range of thicknesses from 90 nm to 500 $\mu$m, with the thickest sample corresponding to bulk HOPG. Every sample was measured at different temperatures ranging from 40 K to 294 K. Compared with a previously reported process for fabricating similar graphite thin film samples, in which a series of wet lithography and dry etching steps were used\cite{Fu2015}, our approach has several advantages. First, we can fabricate very thick samples (on the order of tens of microns) onto the substrates. In contrast, firmly attaching a very thick but narrow graphite strip onto the substrate and maintaining this state during lithographic patterning can be challenging. In this work, we select large area (on the order of millimeters), but generally thin, graphite flakes which can be easily placed on the substrates. Second, by avoiding the etching operation, our method eliminates potential damage to the samples when their lateral geometries are trimmed. \\
 
Thermal properties of graphite flakes were obtained by fitting measured temperature decay curve to a thermal model\cite{Cahill2004,Schmidt2008}. Figures 1c-1f show a typical experimental data set from our TDTR experiment and the corresponding fitting results from a thermal model, demonstrating excellent agreement. The fitting process for such a multi-layer, anisotropic structure is more complicated than for traditional TDTR because there are 4, rather than 2, unknown fitting parameters: the cross-plane thermal conductivity of the graphite flakes, denoted as $k_c$; interface thermal conductance between the Al film and the graphite flake, $G_1$; the anisotropy of the graphite flake, $S$; and the interface thermal conductance between the graphite flake and the sapphire substrate, $G_2$. Further, every different thickness of graphite is a separate sample that may not necessarily have the same value of $G_2$ or anisotropy $S$. To overcome these issues, we performed a multi-step fitting process. We first created a matrix of possible $S$ and $G_2$ values based on general knowledge of the range of these two parameters. Each matrix element consists of an $S$ and a $G_2$ value. Then, we fit the data to the thermal model by fixing $S$ and $G_2$ values from each matrix element and varying $k_c$ and $G_1$ as the fitting parameters. After performing this fitting operation for the whole matrix, a screening was performed by examining the quality of the fit according to a threshold criterion. We recorded the thermal conductivity acquired from each high-quality fit and discarded those from low quality fits. The remaining fitting results are used for calculating the average value and the error bar of the measurement. An example of this procedure is shown in the supplementary information. Following this procedure for each sample and temperature, we extract the thermal conductivity of the graphite flake versus thickness and temperature.\\

Figures 2a and 2b show the results for samples thicker and thinner than 400 nm, respectively. For samples thicker than 400 nm, Figure 2a, the thermal conductivity is largely independent of thickness and exhibits a peak at around 100 K, a common trend that is characteristic of intrinsic phonon-phonon scattering at room temperature and boundary scattering at low temperature. Considering the measured thermal conductivity is independent of thickness, these boundaries must consist of internal grain boundaries. The trend of the curve as well as the peak position are consistent with prior experimental and theoretical works\cite{Ho1972,Fugallo2014}. The measured bulk thermal conductivity at room temperature is between 6.4 W/mK to 7.0 W/mK, in agreement with the accepted value of 6.8 W/mK\cite{Taylor1966}. \\

Samples with thicknesses less than 400 nm, Figure 2b, have a different trend with temperature, slowly increasing rather than exhibiting a peak. This trend is characteristic of samples that have a thermal conductivity limited by boundary scattering. Because the thermal conductivity decreases with decreasing sample thickness, the relevant boundary for these samples is the sample boundary, thereby indicating that c-MFPs are on the order of the sample thickness.  The crossover from the bulk trend to the boundary scattering trend occurs at thickness $\sim$ 200 nm. Our observations are consistent with Fu et al's recent experimental work\cite{Fu2015}, as shown in Figures 2a and 2b.\\

Figures 2c and 2d show the thermal conductivity versus thickness at a series of different temperatures. Owing to the wide range of sample thicknesses, we are able to directly determine the saturation thickness, defined as the thickness above which the sample recovers its bulk thermal conductivity. One can notice that the saturation thickness increases with decreasing temperature, indicating that c-MFPs increase with decreasing temperature, as expected. At temperatures less than 100 K, the saturation thickness reaches around 600 nm and is independent of temperature. This value is much shorter than the thicknesses of the thick samples in this study, which range from microns to hundreds of microns. Therefore, dominant scattering mechanism must be intrinsic scattering by defects or grain boundaries along c-axis, separately confirming the original conclusion obtained from the temperature dependence of the thermal conductivity. \\

Figure 2c also illustrates a comparison with experimental data from Fu et al in a similar range of thicknesses at room temperature\cite{Fu2015}, showing a quite consistent trend. However, due to the limitation of the thickness of the samples used by Fu et al, the maximum of which was 714 nm, the saturation value was only estimated via theoretical extrapolation. Here, we are able to directly measure the saturation thickness and the bulk value. We notice that our measured bulk values are higher than those estimated by Fu et al. One possible reason for this difference is that the etching process required for the 3$\omega$ method may introduce additional defects that scatter phonons, lowering the bulk thermal conductivity. Therefore, maintaining large sample sizes in the basal plan is still important even for measuring cross-plane thermal properties. \\

Our measurements show that scattering from internal grain boundaries largely sets the c-axis thermal conductivity of graphite at low temperatures. To identify the structures responsible for grain boundary scattering, we performed transmission electron microscopy (TEM) studies on our samples. Figure 3a, a bright field (BF) TEM image, clearly shows the existence of some grains of different contrast from the bulk (dark regions), which are separated by a few hundred nanometers on average. Overall, the sample is highly crystalline as confirmed by diffraction patterns of the sample (see SI). These observations are consistent with Park et al's work\cite{Park2010}.\\ 

Figure 3b is a dark field TEM image of the same region in Figure 3a. All the bright regions indicate grains of a specific crystalline orientations, which are distinct from the other areas of the sample. These bright regions can be readily located as corresponding dark regions in the same areas of Figure 3a, which means that the variations of contrast in BF images originates from grains of different crystalline orientations.\\ 

Figure 3c is a magnified BF TEM image on the same sample. Considerable contrast is observed even within the dark regions, indicating differences in crystal structure between the bright background and these regions. High resolution TEM images on the boundary of dark and bright regions, Figure 3d, shows that both areas demonstrate good crystallinity and atomic registry at the interface.  These observations imply that the grains maintain the same c-axis with the rest of the sample but are of different rotational orientations. Because the saturation thickness is on the order of hundreds of nanometers, precisely the spacing of these rotational mismatches, our results indicate that the rotational orientation differences are the structural feature responsible for phonon scattering. \\

With these data, we can qualitatively estimate that the typical phonon c-MFP on the order of hundreds of nanometers at room temperature. However, to obtain quantitative information on the c-MFP spectrum, we need a way to link the measured thermal conductivities to the c-MFP spectrum. We have recently reported a theoretical approach that makes this link\cite{Minnich2012,Zhang2015, Hua2015}, which we now apply here.\\

The cross-plane thermal conductivity reduction in graphite thin films can be attributed to boundary scattering of phonons\cite{Hua2015} in a conceptually similar manner to in-plane thermal conductivity reduction described by Fuchs-Sondheimer theory\cite{Fuchs1938,Sondheimer1952}. In a recent work, we derived the equation that relates the cross-plane thermal conductivity to the film thickness for an isotropic crystal:\\
\begin{equation}
k_c=\int_{0}^{\infty}S(\text{Kn}_{\omega})f(\Lambda_{\omega})d\Lambda_{\omega}=\int_{0}^{\infty} L^{-1} K(\text{Kn}_{\omega})F(\Lambda_{\omega}) d\Lambda_{\omega}
\end{equation}
where $\text{Kn}_{\omega}=\Lambda_{\omega}/L$ is the Knudsen Number, $\Lambda_{\omega}$ is the c-MFP, $L$ denotes the sample thickness along the c-axis, $f(\Lambda_{\omega})$ and $F(\Lambda_{\omega})$ are differential and accumulative c-MFP spectra related by $F(\Lambda_{\omega})=\int_{0}^{\Lambda_{\omega}}f(\Lambda) d\Lambda$, and $S(\text{Kn}_{\omega})$ is the heat flux suppression function that describes the reduction in phonon MFP due to boundary scattering. The kernel $K(\text{Kn}_{\omega})$ is defined as $K(\text{Kn}_{\omega})=-dS(\text{Kn}_{\omega})/d\text{Kn}_{\omega}$. The suppression function was originally derived in Ref. \citenum{Hua2015} for an isotropic crystal. Repeating the derivation for a crystal with an arbitrary dispersion relation, we find
\begin{equation}
S(\text{Kn}_{c,\textbf{k}}) = 1 - \text{Kn}_{c,\textbf{k}}(1-e^{-\frac{1}{\text{Kn}_{c,\textbf{k}}}})
\end{equation}
where $\text{Kn}_{c,\textbf{k}}=\Lambda_{c,\textbf{k}}/L$ is the Knudsen number along the c-axis, defined as the component of the MFP along the c-axis $\Lambda_{c,\textbf{k}}$ normalized by the thickness of the sample $L$. Both $S(\text{Kn}_{c,\textbf{k}})$ and $K(\text{Kn}_{c,\textbf{k}})$ are plotted in the inset of Figure 4a. \\

Using this function, the c-MFP spectra can be recovered from the data by solving an inverse problem as described in Ref. \citenum{Minnich2012}. The original experimental data and the reconstructed c-MFP spectra at different temperatures are shown in Figure 4. At room temperature, the c-MFP spectrum has a minimum and maximum of around 40 nm and 250 nm, respectively. The result is consistent with Yang et al's qualitative estimation that the average of c-MFPs can be well over 100 nm\cite{Yang2014}. As temperature decreases, both bounds increase and the bandwidth increases monotonically. At 40 K, the lower and upper bounds have increased to about 100 nm and 600 nm, respectively.\\

Note that the width of the spectrum is considerably narrower than in common isotropic crystals such as Si\cite{Ward2010,Esfarjani2011}, which has a MFP spectrum spanning at least four orders of magnitude. We attribute the narrow bandwidth to two main factors. First, along the c-axis, graphite is a vdW solid that does not support high frequency phonons. As a result, the range of phonon frequencies that contribute to c-axis heat conduction is considerably narrower than in isotropic crystals. Only phonons whose frequencies are lower than 4 THz can effectively transport along the c-axis\cite{Dolling1962}. The lack of contribution from high frequency phonons imposes a restriction on the minimum possible c-MFP. Second, grain boundaries reduce the maximum c-MFPs to a value comparable to the grain size. Therefore, the c-MFP spectrum is restricted both at small and large c-MFPs. \\

We plot the differential c-MFP spectra at various temperatures in Figure 5a, allowing the change of the bandwidth and peak of the spectrum with temperature to be clearly observed. As temperature decreases, both the upper and lower bounds of the c-MFPs increase. Meanwhile, a small broadening of the bandwidth can be observed.\\ 

To more compactly describe these distributions, we define c-MFP$_{P}$ as the c-MFP at the maximum of the differential distribution. Additionally, we calculate the midpoint MFP, c-MFP$_{mid}$, which indicates the 50\% point of the accumulative distributions in Fig 4. The two parameters, together with lower and upper c-MFP bounds as functions of temperature, are shown in Figure 5b. Although all the four parameters increase with decreasing temperature, the upper bound of c-MFP increases faster than the rest.  The c-MFP$_{P}$ is almost at the first trisection point of between the upper and lower bounds. These results provide a clear indication of the average c-MFP and the bandwidth at a range of temperatures in graphite.\\


Our study of c-MFP spectra of graphite provides important scientific insights as well as guidance for engineering graphitic thermal materials for applications such as graphite nanoplatelet papers \cite{Xiang2011,Veca2009} and thermal interface materials\cite{Tian2013}. In particular, our work shows that the c-axis thermal conductivity is only achieved in graphite films larger than 400 nm at room temperature; thinner films will conduct heat less effectively along the c-axis than would be expected according to the bulk thermal conductivity due to boundary scattering. On the other hand, decreasing the thermal conductivity appears to be quite achievable due to the long c-MFPs. One potential approach to achieve lower thermal conductivity is introducing nanoparticles in the bulk graphite between layers or grains\cite{ Lu2012, Dubrovkin2015}, or, as discussed in this work, introducing additional rotational mismatches of grains to scatter phonons.\\

Scientifically, our work demonstrates that the microscopic properties of thermal phonons can be considerably different in highly anisotropic materials than in isotropic solids. In particular, the c-axis MFP spectrum of graphite is much narrower than in isotropic crystals due to restrictions on both the maximum and minimum c-MFPs. Additionally, our approach of combining thermal measurements over length scales comparable to MFPs with TEM characterization shows considerable promise for studying the microscopic processes governing heat conduction in other anisotropic solids.\\


In conclusion, we have reported the first quantitative measurements of the c-axis MFPs of thermal phonons in graphite. We find that at room temperature, the c-MFPs are on the order of 100-200 nm and are primarily limited by intrinsic phonon-phonon scattering, while at low temperatures boundary scattering from internal grains governs the thermal transport. We also find that the c-MFP spectrum of graphite is unusually narrow compared with that of common isotropic crystals due to the narrow range of phonon frequencies that are supported along the c-axis. Our work provides important insights necessary to understand and engineer heat conduction in graphite and other vdW solids.\\

\section*{Acknowledgements}

The authors are grateful to the Kavli Nanoscience Institute at Caltech for the availability of critical cleanroom facilities and to the Lewis Group at Caltech for use of certain facilities. The authors also thank Matthew H. Sullivan for FIB assistance and Carol M. Garland for TEM assistance. This work was supported by a start-up fund from the California Institute of Technology and by the National Science Foundation under CAREER Grant CBET 1254213.

\section*{Additional Information}
The authors declare no competing financial interests.

\pagebreak

\providecommand{\latin}[1]{#1}
\providecommand*\mcitethebibliography{\thebibliography}
\csname @ifundefined\endcsname{endmcitethebibliography}
  {\let\endmcitethebibliography\endthebibliography}{}

\pagebreak

\begin{figure}[H]
\centering
  \includegraphics[width=0.99\textwidth]{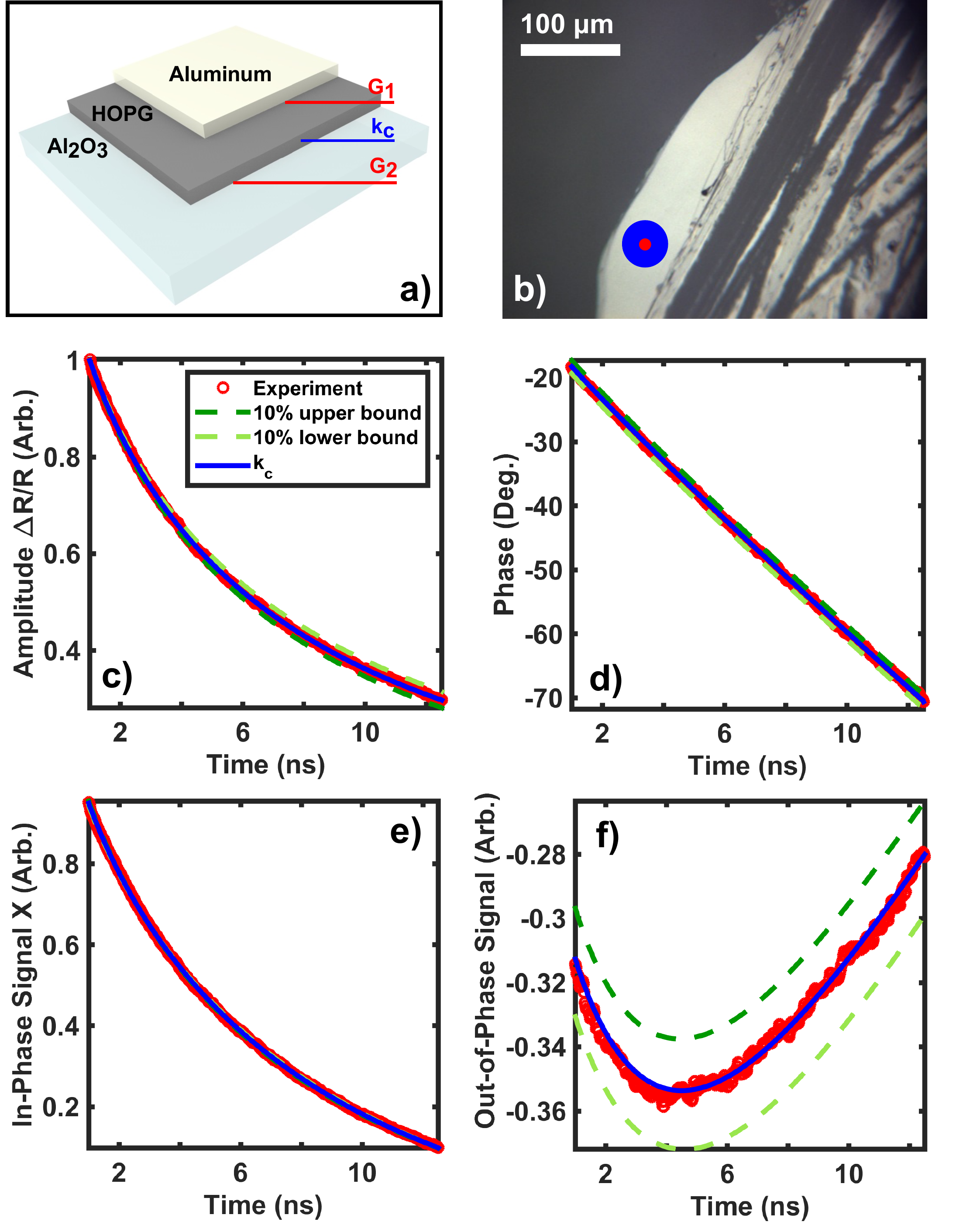}
\end{figure}

\begin{figure}[H]
\centering
  \caption{Typical sample structure, experimental data and their corresponding fitting curves. (a) A three-dimensional cutaway view of a typical sample. (b) An optical image of a graphite sample exfoliated on top of sapphire substrate. The solid blue and red circles indicate typical sizes of pump and probe spots, respectively. The scale bar is 100 $\mu$m. The (c) normalized amplitude, (d) phase, (e) in-phase, and (f) out-of-phase signals from the lock-in amplifier are presented as a function of delay time.  The blue solid lines indicate the result of the fitting using the thermal model, showing an excellent fit. The dark and light green dash lines indicate the cases when $k_c$ is 10 \% higher and lower than the ideal fitting result, respectively.}
\end{figure}

\begin{figure}[H]
\centering
  \includegraphics[width=0.99\textwidth]{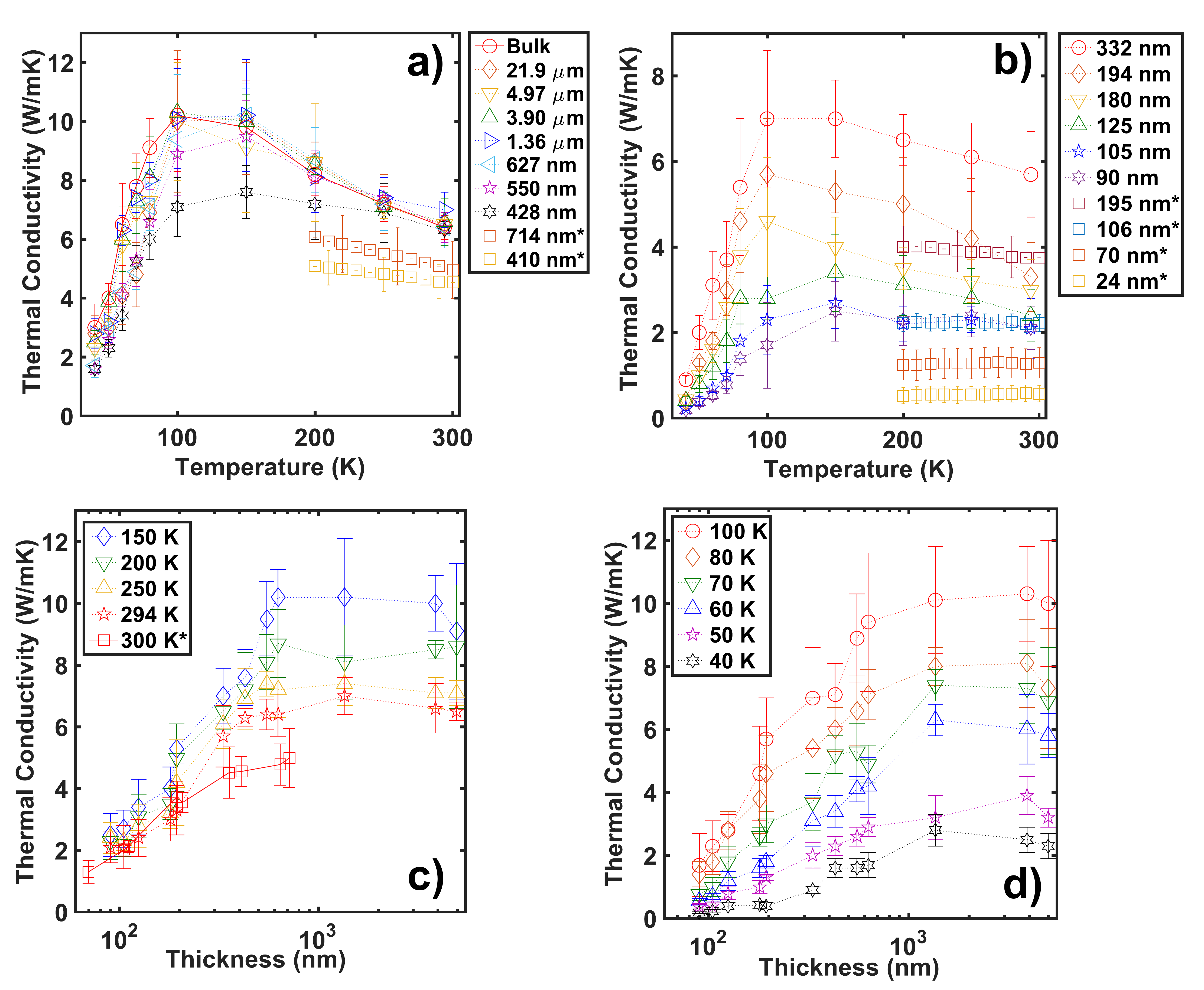}
  \caption{Temperature dependent and thickness dependent thermal conductivities. Temperature dependent data (symbols connected with dotted lines) of samples thicker than 400 nm (a) and thinner than 400 nm (b) . Data from Ref. \citenum{Fu2015} (squares) for the same thickness range is also shown. Similar trends are observed. Thermal conductivity of graphite samples as a function of thickness for temperatures (c) greater than 100 K and (d) less than and including 100 K. Experimental data from Ref. \citenum{Fu2015} at room temperature only (squares) are also shown here for comparison. Both sets of room temperature data present the same trend. ``*" denotes the data from Ref. \citenum{Fu2015}.}
\end{figure}

\begin{figure}[H]
\centering
  \includegraphics[width=0.99\textwidth]{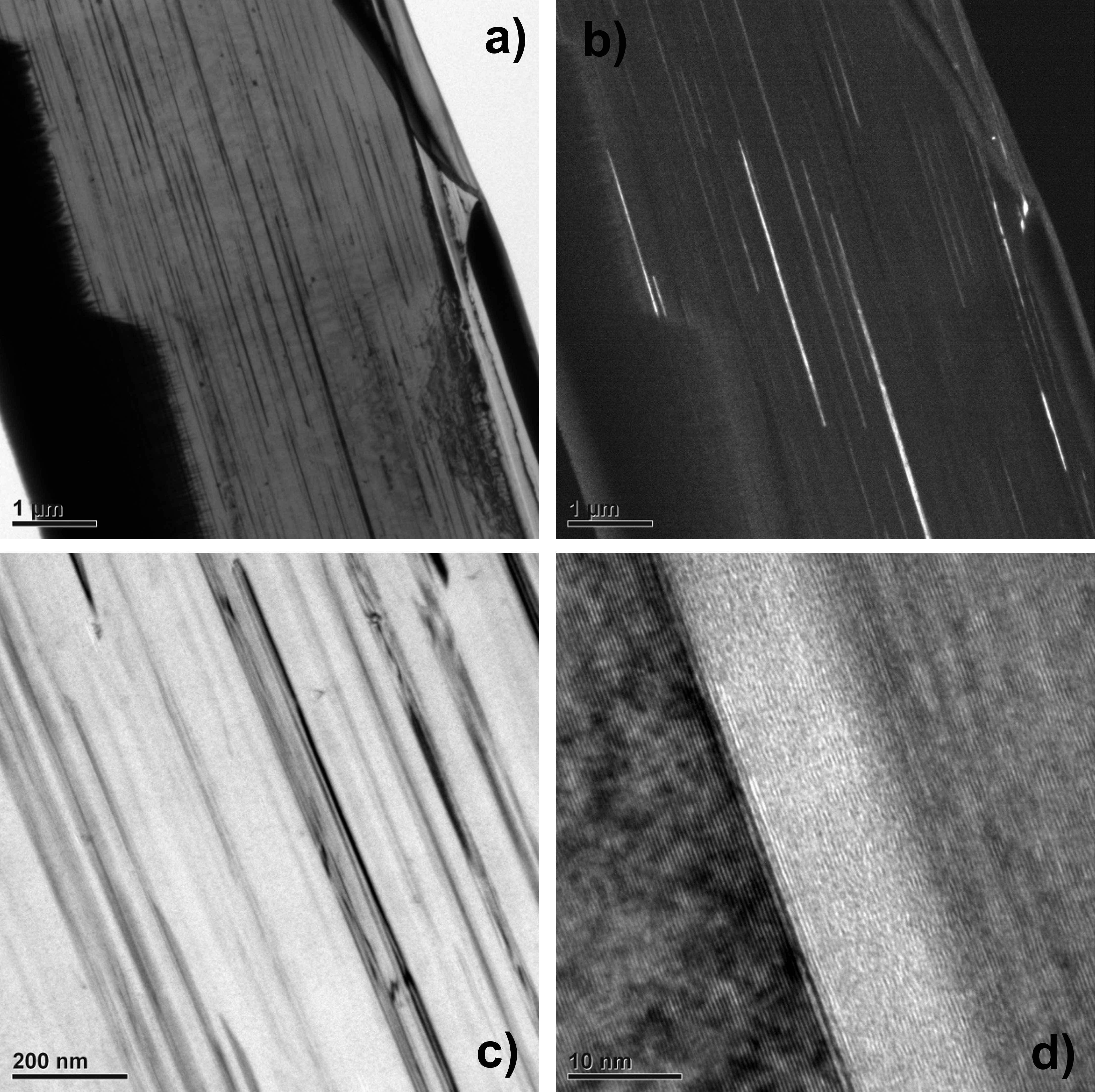}
  \caption{TEM images on the cross section of a bulk graphite sample. Bright field (BF) (a) and dark field (DF) (b) TEM images on the same region of the same sample. Numerous wide dark strips in the BF image can be easily located in the DF image, which are shown as bright strips and indicate grains of a specific orientation distinct from the other areas. (c) A magnified BF TEM image of dark and bright regions. Non-uniformity within dark regions can be identified by the fluctuation in the contrast. (d) High resolution TEM image on the boundary of a dark and a bright region. Both regions demonstrate nice atomic layered structure, which indicates that they share the same c-axis but are of different rotational orientations. Scale bars are 1 $\mu$m (a), 1 $\mu$m (b), 200 nm (c) and 10 nm (d), respectively.}
\end{figure}

\begin{figure}[H]
\centering
  \includegraphics[width=0.99\textwidth]{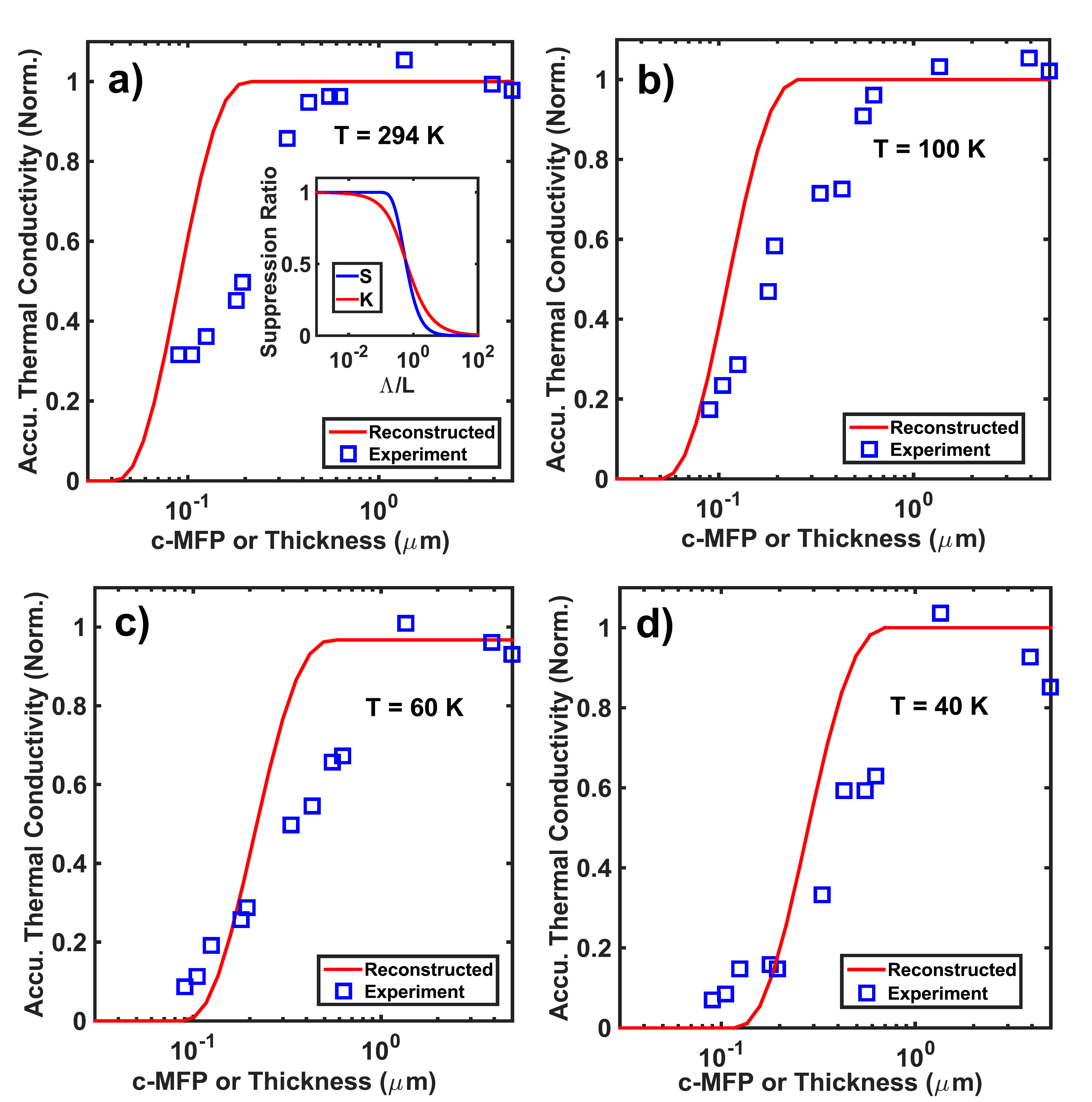}
  \caption{Thickness dependent thermal conductivities and c-MFP spectra reconstruction. Experimental measurements of $k_c$ (blue open squares) and the corresponding reconstructed accumulative thermal conductivity as a function of c-MFP (red lines) in graphite samples at various temperatures: (a) 294 K, (b) 100 K, (c) 60 K and (d) 40 K. As temperature decreases, the c-MFPs increase to a maximum of around 600 nm at 40 K. The x-axis corresponds to thickness for the experimental data and c-MFP for the c-MFP reconstruction. Inset: The suppression function (blue solid line) and the kernel function (red solid line) versus Knudsen number, which are used to perform the reconstruction. }
\end{figure}

\begin{figure}[H]
\centering
  \includegraphics[width=0.99\textwidth]{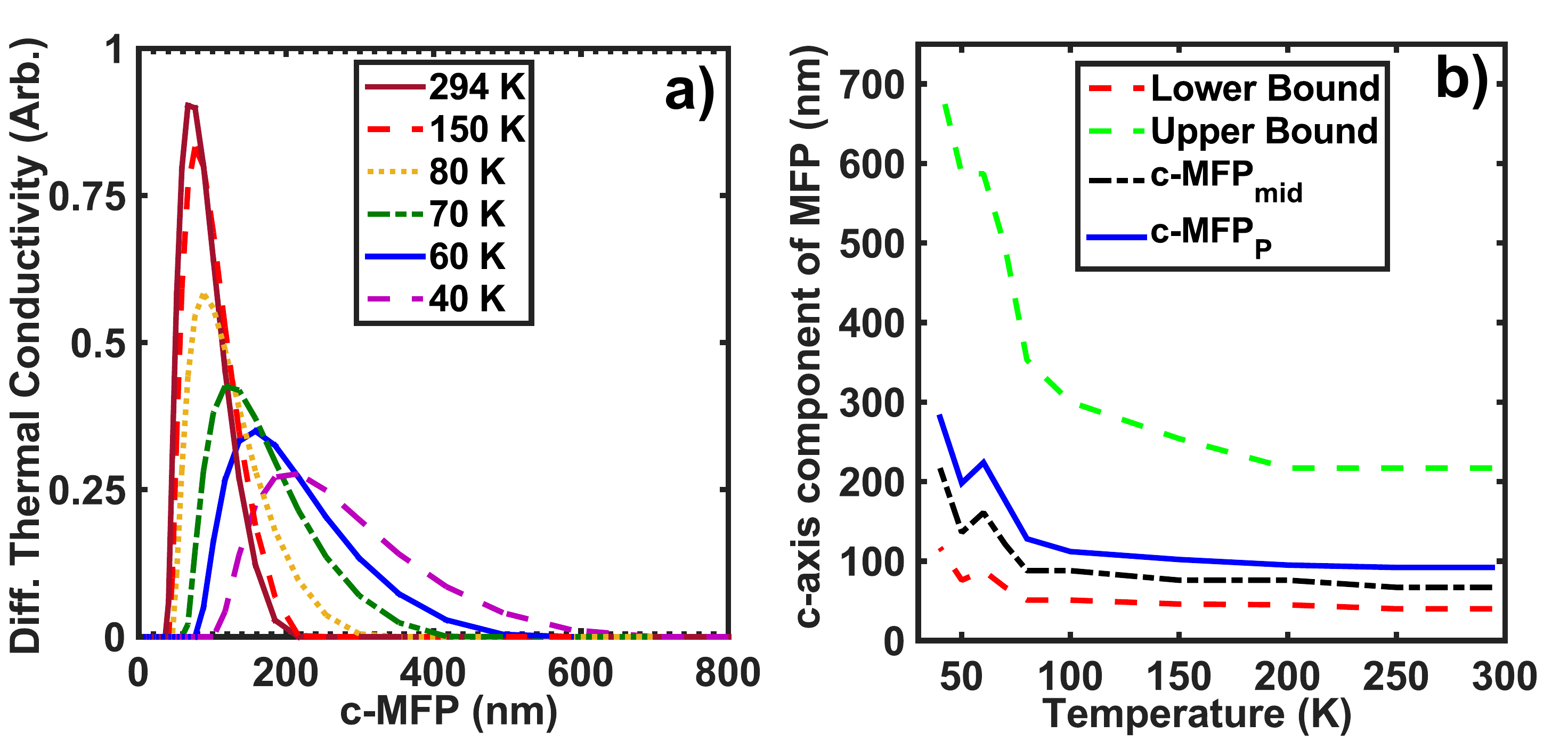}
  \caption{Distributions of c-MFPs in HOPG samples: (a) Differential phonon c-MFP spectra at different temperatures. (b) Upper (green dashed line) and lower (red dashed line) limit of phonon c-MFPs which carry and transport heat. The black dashed line indicates the c-MFP$_{mid}$, or the midpoint c-MFP. The blue solid line indicates the c-MFP$_P$, corresponding to the c-MFP where the maximum of the differential c-MFP spectrum occurs. The c-MFP bandwidth is much narrower than in isotropic crystals.}
\end{figure}

\end{document}